\documentclass[traditabstract]{aa}
\usepackage{graphicx}
\usepackage{natbib}
\usepackage{txfonts}
\bibpunct{(}{)}{;}{a}{}{,}
\begin{document}

  \title{Stellar models with mixing length and $T(\tau)$ relations calibrated on 3D convection simulations}
  
  \author{Maurizio Salaris\inst{1} \and Santi Cassisi\inst{2}}

\institute{Astrophysics Research Institute, 
           Liverpool John Moores University, 
           IC2, Liverpool Science Park, 
           146 Brownlow Hill, 
           Liverpool L3 5RF, UK, \email{M.Salaris@ljmu.ac.uk} 
            \and INAF~$-$~Osservatorio Astronomico di Collurania, Via M. Maggini, I$-$64100 , Teramo, Italy, 
            \email{cassisi@oa-teramo.inaf.it}  
           }

 \abstract{The calculation of the thermal stratification in the superadiabatic layers of stellar models with convective envelopes  
is a long standing problem of stellar astrophysics, and has a major impact on predicted observational properties like radius and 
effective temperature. 
The Mixing Length Theory, almost universally used to model the superadiabatic convective layers, 
contains effectively one free parameter to be calibrated --$\alpha_{\rm ml}$-- whose value controls the resulting 
effective temperature. 
Here we present the first self-consistent stellar evolution models calculated by employing the atmospheric temperature stratification, 
Rosseland opacities, and calibrated 
variable ${\rm \alpha_{\rm ml}}$ (dependent on effective temperature and surface gravity) 
from a large suite of three-dimensional radiation hydrodynamics simulations 
of stellar convective envelopes and atmospheres for solar stellar composition (Trampedach et al.~2013). 
From our calculations (with the same composition of the radiation hydrodynamics simulations),  
we find that the effective temperatures of models with the hydro-calibrated variable $\alpha_{\rm ml}$ (that ranges 
between $\sim$1.6 and $\sim$2.0 in the parameter space covered by the simulations)  
display only minor differences, by at most $\sim$30-50~K, compared to models calculated at constant 
solar $\alpha_{\rm ml}$ (equal to 1.76, as obtained from the same simulations). 
The depth of the convective regions is essentially the same in both cases.
We have also analyzed the role played by the hydro-calibrated $T(\tau)$ relationships in determining the evolution of 
the model effective temperatures, when compared to alternative $T(\tau)$ relationships often used in 
stellar model computations.
The choice  of the $T(\tau)$ can have a larger impact than the use of a variable $\alpha_{\rm ml}$ compared to 
a constant solar value. We found 
that the solar semi-empirical $T(\tau)$ by Vernazza et al.~(1981) provides stellar model effective temperatures 
that agree quite well 
with the results with the hydro-calibrated relationships.
}
\authorrunning{M. Salaris \& S. Cassisi}
\titlerunning{Stellar models with 3D hydro-calibrated convection}
  \maketitle


\section{Introduction}

The almost universally adopted method for calculating superadiabatic convective temperature gradients in  
stellar evolution models is based on the formalism provided by the so-called 
Mixing Length Theory \citep[MLT --][]{bv58}. This formalism is extremely simple; the gas flow is made of 
columns of upward and downward moving convective elements with a characteristic size, the same in all dimensions, 
that cover a fixed mean free path before dissolving. All 
convective elements have the same physical properties at a given distance from the star centre; 
upward moving elements release their excess heat into the surrounding 
gas, and are replaced at their starting point by the downward moving elements, that thermalize
with the surrounding matter, thus perpetuating the cycle. 
The MLT is a \lq{local}\rq\ theory, and the evaluation of all relevant physical and chemical quantities 
are based on the local properties of each 
specific stellar layer, regardless of the extension of the whole convective region.

Both the mean free path and the characteristic size of the convective 
elements are assumed to be same for all convective bubbles, and are assigned the same value 
$\Lambda=\alpha_{\rm ml} H_{\rm p}$, the so-called \lq{mixing length}\rq). Here  $\alpha_{\rm ml}$ is a 
free parameter (assumed to be a constant value within the convective regions and along all 
evolutionary phases), and $H_{\rm p}$ is the local pressure scale height.
There are additional free parameters in the MLT, that are generally fixed a priori (versions of the MLT 
with different choices of these additional parameters will be denoted here as different MLT \lq{flavours}\lq), 
so that practically the only free parameter to be calibrated is $\alpha_{\rm ml}$. 
Stellar evolution calculations for a fixed mass and initial chemical composition but 
with varying $\alpha_{\rm ml}$, produce evolutionary tracks with different $T_{\rm eff}$ (and radius) evolution, whereas 
evolutionary timescales and luminosities are typically unchanged.
On the other hand, the prediction of accurate values of $T_{\rm eff}$ (and radii) 
by evolutionary models and stellar isochrones is paramount, among others, 
to study colour-magnitude diagrams of resolved stellar populations, 
empirical mass-radius relations of eclipsing binary systems, and predict reliable integrated spectra (and colours) 
of unresolved stellar systems (extragalactic clusters and galaxies).

In stellar evolution calculations the value of $\alpha_{\rm ml}$ is usually calibrated 
by reproducing the radius of the Sun at the solar age with an evolutionary solar model.
This solar calibrated $\alpha_{\rm ml}$ is then kept fixed in all evolutionary calculations of stars of different 
masses and chemical compositions. 
The exact numerical value of $\alpha_{\rm ml}$ varies amongst calculations by different authors because 
variations of input physics and choices of the outer boundary conditions affect the predicted 
model radii and $T_{\rm eff}$ values, hence require different $\alpha_{\rm ml}$ values to match the Sun.
Regarding the various MLT flavours, \cite{pvi90} and \cite{sc08} have shown how they provide the same 
$T_{\rm eff}$ evolution, once $\alpha _{\rm ml}$ is appropriately recalibrated on the Sun.

As discussed by, e.g., \cite{vbs96} and \cite{scw02} metal poor 
red giant branch (RGB) models calculated with the solar calibrated 
value of $\alpha _{\rm ml}$ (hereafter $\alpha _{ml, \odot}$) are able to reproduce the  $T_{\rm eff}$  
of samples of RGB stars in Galactic globular clusters, within the error bars on the 
estimated $T_{\rm eff}$.   
On the other hand, several authors find that variations of $\alpha _{\rm ml}$ with respect to $\alpha _{\rm ml, \odot}$ 
are necessary to reproduce --just to give some examples-- the red edge of the RGB in 
a sample of 38 nearby Galactic disc stars with radii determined from interferometry 
\citep{piau11}, the asteroseismically constrained radii of a sample of main sequence (MS) {\sl Kepler} targets    
\citep{m12}, observations of binary MS stars in the Hyades \citep{y06} and the 
$\alpha$ Cen system \citep{y07}, a low-mass pre-MS  
eclipsing binary in Orion \citep{stassun}.

The MLT formalism provides only a very simplified description 
of convection, and there have been several attempts to introduce non-locality in the MLT  
\citep[see, e.g.,][and references therein]{gna93, deng06}. These \lq{refinements}\rq\ of the MLT 
are often complex and introduce additional free parameters to be calibrated. 
The alternative model by \cite{cm91} and \cite{cm92} includes a spectrum of eddy sizes (rather than the one-sized 
convective cells of the MLT) and fixes the scale length of the convective motions to the distance 
to the closest convective boundary.
Recently \cite{pasetto14} have presented a new non-local and time-dependent model based on the 
solution of the Navier-Stokes equations for an incompressible perfect fluid, that does not  
contain any free parameter\footnote{This model has not been implemented yet in any stellar evolution 
computations.}.

An alternative approach to model the superadiabatic layers of convective envelopes is based on the 
computation of realistic multidimensional radiation hydrodynamics (RHD) simulations of atmospheres and convective envelopes
--where convection emerges from first principles-- that cover the 
range of effective temperatures ($T_{\rm eff}$), surface gravities ($g$), and compositions typical of stars with surface convection. 
These simulations have reached nowadays a high level of sophistication 
\citep[see, e.g.,][]{nsa09} and for ease of implementation in stellar evolution codes, 
their results can be used to provide 
an \lq{effective}\rq hydro-calibration of $\alpha _{\rm ml}$, even though RHD simulations 
do not confirm the basic MLT picture of columns of convective cells. 
After early attempts from rather crude two-dimensional (2D) and three-dimensional (3D) simulations \citep[see, e.g.,][]{dv80, lfs92}, 
a first comprehensive RHD calibration of  $\alpha _{\rm ml}$ was presented by \cite{lfs99} and \cite{fls99}. 
These authors found from their simulations that 
the calibrated $\alpha _{\rm ml}$ varies as a function of metallicity, $g$ and 
$T_{\rm eff}$. \cite{fs99} applied this RHD calibration of $\alpha _{\rm ml}$ 
(plus $T(\tau)$ relations computed from the same RHD models, and Rosseland opacities 
consistent with the RHD calculations) to metal poor stellar evolution models 
for Galactic globular cluster stars, and found that 
the resulting isochrones for the relevant age range have only small 
$T_{\rm eff}$ differences along the RGB (of the order of $\sim$50~K)  
with respect to isochrones computed with a solar calibrated value of $\alpha _{\rm ml}$. 

In the last years a number of grids of 3D hydrodynamics simulations of surface convection have been published, and from the point of view 
of stellar model calculations it is very  
important to study whether the results by \cite{fs99} are confirmed or drastically changed.

\cite{tanner:13} and \cite{tanner:13b} have presented a grid of simulations 
employing in the optically thin layers a 3D Eddington solver \citep{t12}. 
Their calculations cover four metallicities (from Z=0.001 to Z=0.04), but just 
a few $T_{\rm eff}$ values at constant log($g$)=4.30  
for each Z, plus a subset of models at varying He mass fraction for Z=0.001 and Z=0.02. 
These authors studied the properties of convection with varying $T_{\rm eff}$ and chemical composition 
in these solar-like envelopes. \cite{tanner:14}  
extracted metallicity-dependent $T(\tau)$ relations from these same simulations, 
and employed them to highlight the critical role these relations play when calibrating 
$\alpha _{\rm ml}$ with stellar evolution models.

\cite{m13} have published a very large 
grid of 3D RHD simulations for a range of chemical compositions. Their grid covers a range of $T_{\rm eff}$  
from 4000 to 7000~K in steps of 500~K, a range of log($g$) from 1.5 to 5.0 in steps of 0.5 dex, 
and metallicity, [Fe/H], from $-$4.0 to +0.5 in steps of 0.5 and 1.0 ~dex. 
These models have been employed by \cite{mwa15} to calibrate 
$\alpha _{\rm ml}$ as function of $g$, $T_{\rm eff}$ and [Fe/H]. They found that $\alpha _{\rm ml}$ depends in a complex way 
on these three parameters, but in general $\alpha _{\rm ml}$ 
decreases towards higher effective temperature, lower surface gravity and higher metallicity. 
So far \cite{mwa15} have provided only fitting 
formulae for $\alpha _{\rm ml}$ but not publicly available 
prescriptions for the boundary conditions and input physics.

Very recently \cite{trampedach13} produced a non-square grid 
of convective atmosphere/envelope 3D RHD simulations for the solar chemical composition.
The grid spans a $T_{\rm eff}$ range from 4200 to 6900~K for MS stars around log($g$)=4.5, 
and from 4300 to 5000~K for red giants with log($g$)=2.2, the lowest surface gravity available. 
In \cite{trampedach14} the horizontal and temporal averages of the 3D simulations were then matched 
to 1D hydrostatic equilibrium, spherically symmetric envelope models to calibrate 
$\alpha _{\rm ml}$ as function of $g$ and $T_{\rm eff}$ 
\citep[see][for details about the calibration procedure]{trampedach14}. 
Moreover, the same RHD simulations have been employed by \cite{t14b} to calculate 
$g$- and $T_{\rm eff}$-dependent $T(\tau)$ relations from temporal and $\tau$ (Rosseland optical depth) 
averaged temperatures of the atmospheric layers. 
\cite{trampedach14} also provide routines to calculate their g- and $T_{\rm eff}$-dependent RHD-calibrated  
$\alpha _{\rm ml}$ together with their computed $T(\tau)$ relations, and Rosseland opacities consistent 
with the opacities used in the RHD simulations.
This enables stellar evolution calculations where boundary conditions, superadiabatic temperature gradient and  
opacities of the convective envelope are consistent with the RHD simulations. It is particularly important  
to use both the RHD-calibrated $\alpha _{\rm ml}$ and $T(\tau)$ relations, because the $T_{\rm eff}$ 
of the stellar evolution calculations depends on both these inputs \citep[see, i.e.,][and references therein]{scw02, tanner:14}.

Thanks to this consistency between RHD simulations and publicly 
available stellar model inputs, we 
present and discuss in this paper the first stellar evolution calculations where 
\cite{trampedach14} 3D RHD-calibration of 
$\alpha _{\rm ml}$ is self-consistently included in the evolutionary code. 
In the same vein as \cite{fs99}, we focus on the effect of the calibrated variable $\alpha _{\rm ml}$ on the model $T_{\rm eff}$,  
compared to the case of calculations with  
fixed $\alpha _{\rm ml, \odot}$ (as determined from the same RHD simulations). 
Self consistency of opacity and boundary conditions is paramount to assess correctly differential effects, given that 
the response of models to variations of $\alpha _{\rm ml}$ depends on their 
$T_{\rm eff}$, that in turn depends  
on the absolute value of $\alpha _{\rm ml}$, opacities and boundary conditions.  
We also address the role played by the RHD calibrated $T(\tau)$ relations in the determination of the model 
$T_{\rm eff}$ when compared to other widely used relations.

Section~2 describes briefly the relevant input physics of the models, while Sect.~3 presents and compares the resulting evolutionary 
tracks. A summary and discussion close the paper.

\section{Input physics}
 
All stellar evolution calculations presented here have been performed with the BaSTI 
(a BAg of Stellar Tracks and Isochrones) code \citep{basti}, for 
a chemical composition with 
$Y$=0.245, $Z$=0.018 and the \cite{gn93} metal mixture, consistent with the chemical composition of the RHD simulations. 
Atomic diffusion was switched off in these calculations, and convective core overshooting was included when appropriate 
\citep[see][for details]{basti}. 

We employed the Rosseland opacities provided by \cite{trampedach14}. The  
low temperature opacities (for log($T$)$<$4.5) are very close to \cite{faa05} with the exception of 
the region with log($T$)$<$3.5, where \cite{faa05} 
calculations are higher because of the inclusion of the effect of water molecules \citep[see][]{t14b}. 
Opacity Project \citep{op} calculations are used for log($T$)$\geq$4.5.
As for the surface boundary conditions, we employed the $T(\tau)$ relations computed  
by \cite{t14b}, who provided a routine that calculates, for a given $T_{\rm eff}$ and surface gravity, 
the appropriate generalized Hopf functions $q(\tau)$, related to the $T(\tau)$ relation by 

\begin{equation}
q(\tau ) = \frac{4}{3}\left(\frac{T(\tau )}{T_{\rm eff}}\right)^4-\tau 
\end{equation}

In our calculations we fixed to $\tau_{\rm tr}$=2/3 the transition from the $T(\tau)$ integration of the atmospheric layers 
(with $\tau$ as independent variable)  
to the integration of the full system of stellar structure equations. 
As mentioned by 
\cite{t14b} and \cite{trampedach14}, these RHD-based $T(\tau)$ relations can be employed to model also  
the convective layers in the optically thick part of the envelope, together with modified 
expressions for the temperature gradients in the superadiabatic regions, and an appropriate 
rescaling of $\tau$ \citep[see][for details]{t14b}. 
In our calculations we have compared the radiative ($\nabla_{\rm rad}$)
and superadiabatic ($\nabla$ --obtained with the RHD calibrated values of $\alpha _{\rm ml}$) temperature gradients 
determined along the upper part of the convective envelopes from the standard stellar structure equations and MLT
(down to $\tau$=100, the upper limit for the routine calculating the Hopf functions), with $\nabla_{\rm rad}$ and 
$\nabla$ calculated according to Eqs.35 and 36 of \cite{t14b}, respectively. 
We have found that in all our models the differences between these two sets of gradients are 
much less than 1\% between $\tau_{\rm tr}$ and $\tau=100$.

We have calculated also test models by  
changing $\tau_{\rm tr}$ between 2/3 and 5 \citep[with the appropriate rescaling of $\tau$ if convection appears 
in the atmosphere integration, see][]{t14b}, and obtained identical evolutionary tracks in each case.

Regarding the value of $\alpha _{\rm ml}$, 
the same routine for the Hopf functions provides also the RHD calibrated value 
of $\alpha _{\rm ml}$ \citep[for the][flavour of the MLT]{bv58} 
for a given $T_{\rm eff}$ and surface gravity, that we employed in our calculations. 
Uncertainties in the calibrated $\alpha _{\rm ml}$ values are of the order of $\pm 0.02-0.03$ \citep[see Table~1 of][]{trampedach14}
\footnote{We did not include any turbulent pressure in the convective envelope, as the $\alpha _{\rm ml}$ RHD 
calibration was performed in a way that works for standard 
stellar evolution models without this extra contribution to the pressure 
\citep[R. Trampedach private communication, see also Sect.4 from][]{trampedach14}
}.

The only difference in terms of input physics between the atmosphere/envelope RHD calculations 
and our models is the equation of state (EOS). The RHD simulations employed the MHD \citep{mhd} EOS, that is not the 
same EOS used in the BaSTI calculations \citep[see][]{basti}. 
To check whether this can cause major differences in the models, we have calculated envelope models for the same 
g-$T_{\rm eff}$ pairs of the RHD simulations, including the RHD-calibrated $\alpha _{\rm ml}$,  $T(\tau)$ relations, 
and RHD opacities. We have then compared the resulting depths of the convection zones ($d_{\rm CZ}$, in units of stellar radius) with 
what obtained by \cite{trampedach14} from their RHD-calibrated 1D envelope models, that used the same input physics (including EOS) 
of the RHD calculations \citep[see Table~1 of][]{trampedach14}. We found random (non systematic) differences 
of $d_{\rm CZ}$ by at most just 2-3\% compared to \cite{trampedach14} results.

\section{Model comparisons}

\begin{figure}
\centering
\includegraphics[scale=.4500]{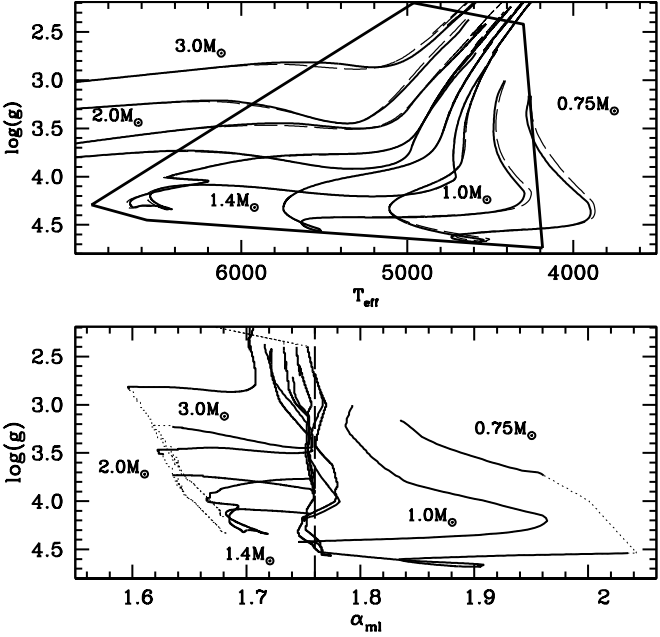}
\caption{Stellar evolution tracks in the g-$T_{\rm eff}$ diagram for the labelled masses. 
The region enclosed by the thick black boundary is the g-$T_{\rm eff}$ 
range covered by the RHD simulations.
Thick solid lines denote fully consistent calculations with the RHD calibrated 
variable $\alpha _{\rm ml}$ and $T(\tau)$ relationships. 
The lower panel displays the evolution of $\alpha _{\rm ml}$ along each track. 
The dotted portion of each sequence denotes the region where the $\alpha _{\rm ml}$ values are extrapolated.
Dashed lines in the upper panel display tracks calculated with a constant $\alpha _{\rm ml}$=$\alpha _{ml, \odot}$ and the 
calibrated $T(\tau)$ relationships.
}
\label{tracks_full}
\end{figure}

As mentioned in the introduction, 
the RHD simulations cover a non-square region in the g-$T_{\rm eff}$ diagram, as displayed in Fig.~\ref{tracks_full} 
(the region enclosed by thick solid lines), 
ranging from 4200 to 6900~K on the MS, and from 4300 to 5000~K for RGB stars with log($g$)=2.2. 
The MLT calibration results in an $\alpha _{\rm ml}$  varying from 1.6 for the warmest dwarfs, with a thin convective envelope, 
up to 2.05 for the coolest dwarfs in the grid. 
In between there is a triangular plateau of $\alpha _{\rm ml}\sim$1.76, where the Sun is located. The RHD simulation for the 
Sun provides  $\alpha _{\rm ml}\sim$1.76$\pm$0.03.
The top panel of Fig.~\ref{tracks_full} displays the results of our   
evolutionary model calculations in the g-$T_{\rm eff}$ diagram (from the pre-MS to the lower RGB), 
for masses M=0.75, 1.0, 1.4, 2.0 and 3.0${\rm M_{\odot}}$ respectively, that cover the full 
domain of the RHD simulations. 

The thick solid lines denote the reference set of models, calculated with the varying $\alpha _{\rm ml}$ calibration, and 
the run of $\alpha _{\rm ml}$ along each individual track is displayed in the lower panel. The dotted part of 
each sequence in this panel denotes the  $\alpha _{\rm ml}$ values extrapolated by the calibration routine, when the models 
are outside the region covered by the simulations but still retain a convective region. 
This happens along the pre-MS evolution of the M=0.85${\rm M_{\odot}}$ 
track, and for the 2.0 and 3.0${\rm M_{\odot}}$ calculations along the subgiant phase. 
Apart from the pre-MS stages of the two lowest mass models, the evolution of 
$\alpha _{\rm ml}$ spans a narrow range of values, between $\sim$1.6 and $\sim$1.8.

To study the significance of the variation of $\alpha _{\rm ml}$ for the model $T_{\rm eff}$, we have 
calculated evolutionary models for the same masses, this time keeping $\alpha _{\rm ml}$=$\alpha _{ml, \odot}$=1.76 
along the whole evolution. The results are also displayed in Fig.~\ref{tracks_full}. 

A comparison of the two sets of tracks clearly shows that the effect of a varying $\alpha _{\rm ml}$ 
is almost negligible. The largest differences are of 
only $\sim$30~K along the RGB phase of the 3.0${\rm M_{\odot}}$ track (solar $\alpha _{\rm }$ tracks being hotter 
because of a higher $\alpha _{\rm ml}$ value compared to the calibration) 
and $\sim$50~K at the bottom of the Hayashi track of the 1.0${\rm M_{\odot}}$ track (solar $\alpha _{\rm ml}$ tracks being cooler, 
because 
of a lower  $\alpha _{\rm ml}$). In all other cases differences are smaller, and often equal to almost zero.

For all stellar masses we found that the mass fraction of He dredged to the surface by the first dredge up 
--that depends on the maximum depth of the convective envelope at the beginning of the RGB phase-- 
is the same within 0.001, between constant $\alpha _{ml, \odot}$ and variable $\alpha _{\rm ml}$ models. 
We have then compared the luminosity of the RGB bump --that also depends on the 
maximum depth of the convective envelope at the first dredge up \citep[see, e.g.,][and references therein]{cs:97,cs13}--  
for the 0.75 and 1.0${\rm M_{\odot}}$ models. We found that the luminosity is unchanged between models with 
constant and variable $\alpha _{\rm ml}$.
This reflect the fact that the depth of the convective envelope is the same between the two sets of models throughout 
the MS phase to the RGB, until the end of the calculations. Small differences appear only during the pre-MS.
To this purpose, for the 1.0${\rm M_{\odot}}$ models we have additionally checked the surface Li abundance 
that survives the pre-MS depletion. 
This provides information about the evolution of the lower boundary of the convective envelope during this phase, 
where according to Fig.~\ref{tracks_full}, $\alpha _{\rm ml}$ shows the largest difference from $\alpha _{ml, \odot}$. 
We found that models with constant $\alpha _{ml, \odot}$ display after the pre-MS a Li abundance just 9\% higher 
than the reference results.

\begin{figure}
\centering
\includegraphics[scale=.4500]{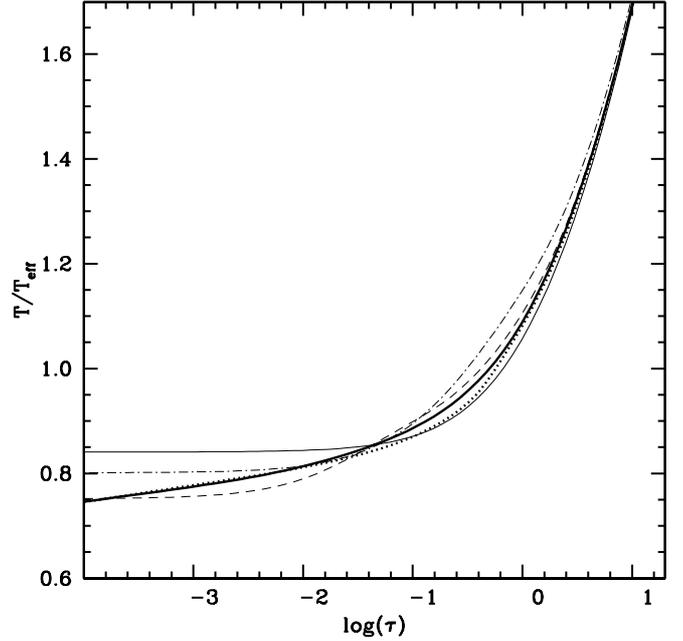}
\caption{Comparison of the ratio $T$/$T_{\rm eff}$ as a function of the optical depth $\tau$ as predicted 
by different $T(\tau)$ relationships. The thick solid and dotted lines display the RHD-calibrated 
relationships for $T_{\rm eff}$=4500~K and 6000~K (log($g$)=3.5 in both cases), respectively. The thin solid, dash-dotted and dashed lines 
show the ratios obtained from the Eddington, KS and ${\rm VAL_c}$ $T(\tau)$ relationships, respectively.
}
\label{ttau_comp}
\end{figure}

We have then analyzed the role played by the $T(\tau)$ relationships computed from the RHD calculations in determining the model 
$T_{\rm eff}$ \citep[the role played by the boundary conditions in determining the $T_{\rm eff}$ of stellar models 
is especially crucial for very-low-mass stars, see i.e.][]{aha97, bcc98}.  
Figure~\ref{ttau_comp} displays the ratio T/$T_{\rm eff}$ as a function of $\tau$ 
as predicted by the calibrated relationships for atmospheres/envelopes with $T_{\rm eff}$=4500~K and 6000~K (log($g$)=3.5). 
The RHD calibrated $T(\tau)$ relationships 
contain values for the Hopf function that vary with $\tau$ and $T_{\rm eff}$ (also with g, to a lesser degree). 
The variation with  $T_{\rm eff}$ is obvious from the figure.
For these two temperatures the largest differences appear at the layers with $\tau$ between $\sim$0.1 and $\sim -$1, where stellar 
model calculations 
usually fix the transition from the atmosphere to the interior.
The same Figure~\ref{ttau_comp} also displays the results for the traditional Eddington approximation to the grey atmosphere,  
and the solar semi-empirical $T(\tau)$ relationships by 
Krishna-Swamy \citep[KS --][]{ks} and \cite{vernazza:81}  --their Model C for the quiet sun, 
hereinafter ${\rm VAL_c T(\tau)}$. 
In these latter cases the ratio T/$T_{\rm eff}$ does not depend on $T_{\rm eff}$.  
It is easy to notice that around the photospheric layers the RHD relationships 
are in between the Eddington grey and ${\rm VAL_c T(\tau)}$. The most discrepant relationship is the KS one.

\begin{figure}
\centering
\includegraphics[scale=.4500]{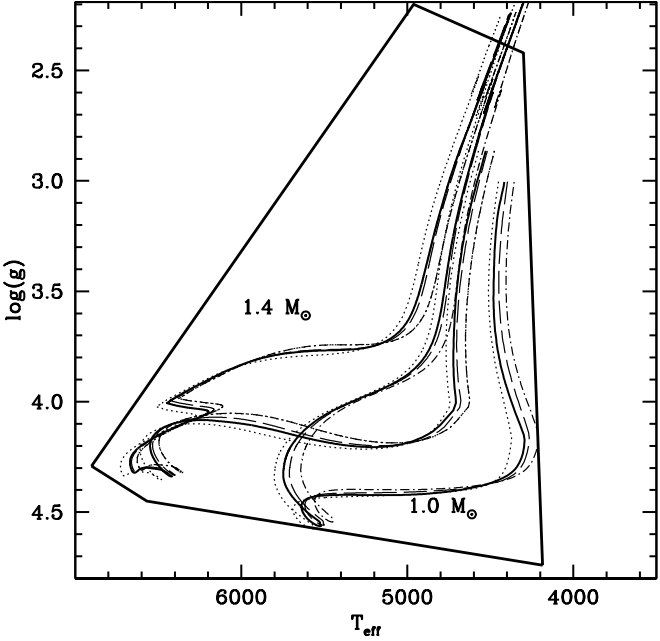}
\caption{As the upper panel of Fig.~\ref{tracks_full} 
but for evolutionary tracks with masses equal to 
1.0 and 1.4${\rm M_{\odot}}$ respectively. Thick solid lines denote 
fully consistent calculations with the RHD calibrated variable $\alpha _{\rm ml}$ and $T(\tau)$ relationships. 
Dotted, dash-dotted and dashed lines display tracks calculated with constant $\alpha _{ml, \odot}$ and the Eddington, 
KS and ${\rm VAL_c}$  T($\tau$), respectively.
}
\label{ttau}
\end{figure}

Figure~\ref{ttau} displays the results of evolutionary calculations with this set of $T(\tau)$ relationships. 
We compare here the reference set of models 
for 1.0 and 1.4${\rm M_{\odot}}$ and varying  $\alpha _{\rm ml}$, with constant $\alpha _{ml, \odot}$ models calculated 
using the Eddington, KS and ${\rm VAL_c}$  $T(\tau)$ relationships  
(we set $\tau_{tr}$=2/3 also for these calculations). 

Differences between these new sets of models at constant $\alpha _{ml, \odot}$ and the reference calculations 
are larger than the case of Fig.~\ref{tracks_full} 
because of the differences with the RHD $T(\tau)$ relationships.
On the whole, the  ${\rm VAL_c}$ $T(\tau)$ (coupled to the RHD calibrated $\alpha _{ml, \odot}$) gives  
the closest match to the self-consistent reference calculations. 
The largest differences appear for the 1.0${\rm M_{\odot}}$ along the pre-MS; models 
calculated with the ${\rm VAL_c}$ relation are $\sim$50~K cooler, approximately the same as  
the case of $\alpha _{ml, \odot}$ and the RHD T($\tau$). For the same  1.0${\rm M_{\odot}}$ track the 
$T_{\rm eff}$ differences along the RGB are at most equal to 10~K, and at most $\sim$40~K along the MS. 
Differences for the 1.4${\rm M_{\odot}}$ track are smaller.

As for the Eddington T($\tau$), the resulting tracks are generally hotter than the self-consistent RHD-based calculations. 
The largest differences amount to $\sim$40-50~K along MS and RGB of the two tracks. Comparisons with Fig.~\ref{tracks_full}   
show that, from the point of view of the resulting model $T_{\rm eff}$, the Eddington $T(\tau)$ differs more --albeit by not much-- 
than the ${\rm VAL_c}$ one from the RHD-calibrated relations.

The worse agreement is found with the KS T($\tau$). For both masses the RGB is systematically cooler by $\sim$70~K, 
and the pre-MS by $\sim$80-100~K, whilst the MS of the 1.0${\rm M_{\odot}}$ calculations is cooler by $\sim$120~K, 
a difference reduced 
to $\sim$50~K along the MS of the 1.4${\rm M_{\odot}}$ track. 

In case of the 1.0${\rm M_{\odot}}$ models we checked again the luminosity level of the RGB bump, 
the surface He mass fraction after the first dredge up and 
the amount of surface Li after the pre-MS depletion. 
For all these three $T(\tau)$ relations the evolutionary models display a RGB bump luminosity within 
$\Delta {\rm (L/L_{\odot})}<$0.01, and a post-dredge up He abundance within 0.001 
of the fully consistent result.
In all these calculations the convective envelope has the same depth throughout 
the MS phase and along the RGB. 
Differences along the pre-MS are highlighted by the amount of Li depletion during this phase.
We found that models calculated with the Eddington $T(\tau)$ have 9\% less Li after the pre-MS, compared to 
the reference calculations. Comparing this number with just the effect of a constant $\alpha _{ml, \odot}$ discussed above, 
we derive that the use of this $T(\tau)$ decreases the surface Li abundance by $\sim$20\% 
compared to the use of the Hopf functions determined from the RHD simulations. 
In case of the ${\rm VAL_c}$ T($\tau$), the net effect is to increase the surface Li after the pre-MS 
by $\sim$10\% compared to the RHD T($\tau$), and also  
the KS relation causes a similar increase by $\sim$11\%.

\subsection{The standard solar model}

We close our analysis by discussing the implications of the RHD results and the choice of the T($\tau$) relation, 
on the calibration of the standard solar model. 
As well known, in stellar evolution it is customary to fix the value of $\alpha _{ml, \odot}$ (and the initial solar He and metal mass fractions) 
by calculating a 1$M_{\odot}$ stellar model that matches the solar bolometric luminosity and radius at the age of the Sun, with the additional 
constraint of reproducing the present metal to hydrogen mass fraction $Z/X$ ratio \citep[see, e.g.,][for details]{basti}.  
The accuracy of the derived solar model can then be tested against helioseismic estimates of the depth of the 
convective envelope and the surface He mass fraction. 
It is also well established that solar models without microscopic diffusion cannot properly account
for some helioseismic constraints, hence solar models are routinely calculated by including microscopic diffusion 
of He and metals \citep[see, e.g.,][for a discussion and references]{basti}.

We have first calculated a standard solar model (with the same input physics and solar metal distribution 
of the calculations discussed above) employing both the variable $\alpha _{\rm ml}$ and the T($\tau$) relations from the RHD results. 
Given that microscopic diffusion decreases with time the surface chemical abundances of the model, the initial solar $Z$ (and $Y$) 
need to be higher than the present one. This means that we had to employ the RHD results also for chemical compositions not exactly the 
same as the composition of the RHD simulations.

We found that it is necessary to rescale the RHD $\alpha _{\rm ml}$ calibration by a factor of just 1.034  
to reproduce the solar radius. This implies $\alpha _{\rm ml, \odot}$=1.82, extremely close, within the error, to the RHD value  
$\alpha _{\rm ml, \odot}$=1.76$\pm$0.01(range)$\pm$0.03(calibration uncertainty) obtained by \citep{trampedach14}. 
Given our previous results, it is also obvious that a solar model calibration with fixed $\alpha _{\rm ml}$ and 
the RHD T($\tau$) relations provides the same $\alpha _{\rm ml, \odot}$=1.82. 

Solar calibrations with the KS, ${\rm VAL_c}$ and Eddington T($\tau$) relations have provided 
$\alpha _{\rm ml, \odot}$=2.11, 1.90 and 1.69, respectively. 

In all these calibrated solar models 
the initial solar He mass fraction ($Y_{\rm ini, \odot}\sim 0.274$) and metallicity (($Z_{\rm ini, \odot}\sim 0.0199$) 
are essentially the same, as expected. Also, the model present He mass fraction in the envelope ($Y_{\odot}= 0.244$)  
and the depth of the convection zone ($d_{\rm CZ}=0.286 R_{\odot}$) are the same 
for all calibrations and in agreement with the helioseismic values \citep[see, e.g.,][]{d95, ba97}.

If we take $\alpha _{\rm ml, \odot}$ obtained with the 
RHD T($\tau$) as a reference, the lower value obtained with the Eddington T($\tau$), and the larger values obtained 
with both the ${\rm VAL_c}$ and KS relations (in increasing order) are fully consistent with the results of Fig.~\ref{ttau}. 
In that figure ${\rm VAL_c}$ and KS T($\tau$) MS models are increasingly hotter than the reference RHD calculations (hence increasingly larger 
$\alpha _{\rm ml}$ values are required to match the reference MS) 
whereas the use of the Eddington T($\tau$) produces models cooler than the reference MS (hence lower 
$\alpha _{\rm ml}$ value are needed to match the reference MS).

\section{Summary and discussion}

We have presented the first self-consistent stellar evolution calculations that employ the variable 
$\alpha _{\rm ml}$ and $T(\tau)$ by \cite{trampedach14}, based on their 3D RHD simulations.
Our set of evolutionary tracks for different masses and the same chemical composition 
(plus consistent Rosseland opacities) of the RHD simulations, cover approximately the entire 
g-$T_{\rm eff}$ parameter space of the 3D atmosphere/envelope calculations. 

We found that, from the point of view of the predicted $T_{\rm eff}$ (plus the depth of the convective 
envelopes and amount of pre-MS Li depletion), 
models calculated with constant RHD calibrated $\alpha _{\rm ml, \odot}=1.76$ are very close to,  
and often indistinguishable from, the models with variable $\alpha _{\rm ml}$. Maximum differences are 
at most $\sim$30-50~K. This result is similar to the conclusions by \cite{fs99}, based on 
2D RHD simulations at low metallicities.
 
At first sight this may appear surprising, given that the full range of $\alpha _{\rm ml}$ spanned by the 
RHD calibration is between $\sim$1.6 and $\sim$2.0. However, one has to take into account the following points:

\begin{enumerate} 
 
\item{The derivative $\Delta T_{\rm eff}/\Delta \alpha _{\rm ml}$ for stellar evolution models depends 
on the absolute value of $\alpha _{\rm ml}$, and decreases when $\alpha _{\rm ml}$ increases).
}

\item{The variation of $\Delta T_{\rm eff}$ for a given $\Delta \alpha _{\rm ml}$ depends also on the 
mass extension of the convective and superadiabatic regions.
}

\end{enumerate}

It is therefore clear that the decrease of $\alpha _{\rm ml}$ with increasing $T_{\rm eff}$ cannot have a major effect 
because of the thin convective (and superadiabatic) layers of models crossing this region of the g-$T_{\rm eff}$ diagram. 
The large variations $\Delta \alpha _{\rm ml}\sim$0.2 
(see lower panel of Fig.~\ref{tracks_full}) along the lower Hayashi track of the 1.0${\rm M_{\odot}}$ model is also not 
very significant ($\sim$50~K) because of the decreased extension of the surface convection and the reduced 
$\Delta T_{\rm eff}/\Delta \alpha _{\rm ml}$ at higher $\alpha _{\rm ml}$.

It is however very important to remark here that the detailed structure of the superadiabatic convective 
regions is not suitably reproduced either by 
$\alpha _{ml, \odot}$ or by a variable $\alpha _{\rm ml}$, and that the full results from RHD models need to be employed whenever 
a detailed description of the properties of these layers is needed. 

To some degree the role played by the RHD $T(\tau)$ is more significant. 
Constant $\alpha _{\rm ml, \odot}=1.76$ stellar evolution models become systematically cooler by up to $\sim$100~K  
along the MS, pre-MS and RGB 
when the widely used KS $T(\tau)$ relation is used, compared to the self consistent RHD-calibrated calculations.
Evolutionary tracks obtained employing the Eddington and  ${\rm VAL_c}$ $T(\tau)$ 
relationships are much less discrepant, and stay within $\sim$50~K of the self-consistent calculations. 
Even from the point of view of pre-MS Li depletion the ${\rm VAL_c}$ $T(\tau)$ causes only minor differences, of the order of 10\%. 
A similar difference is found with the KS relation, 
whilst a larger effect of $\sim$20\% is found with the Eddington T($\tau$).  

An extension of \cite{trampedach13} simulations to different metallicities is necessary to extend this study 
and test the significance of the $\alpha _{\rm ml}$ variability (and Hopf functions) over a larger parameter space. 
The 3D RHD simulations by \cite{m13} and the $\alpha _{\rm ml}$ calibration by \cite{mwa15} cover a large metallicity range, 
and predict an increase of $\alpha _{\rm ml}$ with decreasing metal content, but  
at the moment it is not possible to properly include this calibration in stellar evolution calculations, for the lack 
of available $T(\tau)$ relations and input physics consistent with the RHD simulations.
At solar metallicity the general behaviour of $\alpha _{\rm ml}$  with $g$ and $T_{\rm eff}$ seems to be qualitatively similar 
to \cite{trampedach14} calibration. However, the range of $\alpha _{\rm ml}$ is shifted to higher values compared 
to \cite{trampedach14} results, due probably to different input physics and the different adopted solar chemical composition.

\begin{acknowledgements}
We are grateful to R. Trampedach for clarifications about his results, and for making available  
the routine to calculate Rosseland opacities. 
We also thank J. Christensen-Dalsgaard for interesting discussions on this topic, and the anonymous referee 
for comments that have improved the presentation of our results.
SC warmly acknowledges financial support from PRIN-INAF2014 (PI: S. Cassisi) 
\end{acknowledgements}

\bibliographystyle{aa}

\end{document}